\documentclass[twocolumn,showpacs,preprintnumbers,superscriptaddress]{revtex4}
\usepackage{graphicx,bm,times,url}
\usepackage{amsmath}
\graphicspath{{./fig/}{./png/}}
\newcommand{\BoldVec}[1]{\mathchoice%
  {\mbox{\boldmath $\displaystyle     #1$}}%
  {\mbox{\boldmath $\textstyle        #1$}}%
  {\mbox{\boldmath $\scriptstyle      #1$}}%
  {\mbox{\boldmath $\scriptscriptstyle#1$}}%
}
\newcommand{\EQ}{\begin{equation}}
\newcommand{\EN}{\end{equation}}
\newcommand{\EQA}{\begin{eqnarray}}
\newcommand{\ENA}{\end{eqnarray}}

\newcommand{\Eq}[1]{Eq.~(\ref{#1})}

\newcommand{\Eqss}[2]{Eqs.~(\ref{#1})--(\ref{#2})}

\newcommand{\Fig}[1]{Fig.~\ref{#1}}

\newcommand{\Figs}[2]{Figs.~\ref{#1} and \ref{#2}}

\newcommand{\nf}{\mbox{$n_{\rm f}$}}
\newcommand{\napp}{\mbox{$n_{\rm app}$}}
\newcommand{\Hm}{\mbox{$H_{\rm M}$}}
\newcommand{\xx}{\BoldVec{x}{}}

\newcommand{\UU}{\BoldVec{U} {}}

\newcommand{\BB}{\BoldVec{B} {}}

\newcommand{\AAA}{\BoldVec{A} {}}

\newcommand{\JJ}{\BoldVec{J} {}}

\newcommand{\FF}{\BoldVec{F} {}}

\newcommand{\nab}{\BoldVec{\nabla} {}}

\newcommand{\SSSS}{\bm{\mathsf{S}}}

\newcommand{\DD}{{\rm D} {}}

\def\Pm{\mbox{\rm Pr}_{\rm M}}

\def\Lu{\mbox{\rm Lu}}

\begin{document}
\preprint{NORDITA 2011-26}

\title{Decay of helical and non-helical magnetic knots}
\author{Simon Candelaresi}
\affiliation{NORDITA, AlbaNova University Center, Roslagstullsbacken 23,
SE-10691 Stockholm, Sweden}
\affiliation{Department of Astronomy,
Stockholm University, SE 10691 Stockholm, Sweden}

\author{Axel Brandenburg}
\affiliation{NORDITA, AlbaNova University Center, Roslagstullsbacken 23,
SE-10691 Stockholm, Sweden}
\affiliation{Department of Astronomy,
Stockholm University, SE 10691 Stockholm, Sweden}

\date{\today,~ $ $Revision: 1.136 $ $}
\begin{abstract}
We present calculations of the relaxation of magnetic field structures
that have the shape of particular knots and links.
A set of helical magnetic flux
configurations is considered, which we call $n$-foil knots of which the trefoil
knot is the most primitive member.
We also consider two nonhelical knots; namely, the Borromean rings as well
as a single interlocked flux rope that also serves as the logo of the
Inter-University Centre for Astronomy and Astrophysics in Pune, India.
The field decay characteristics of both configurations is
investigated and compared with
previous calculations of helical and nonhelical triple-ring configurations.
Unlike earlier nonhelical configurations, the present ones cannot
trivially be reduced via flux annihilation to a single ring.
For the $n$-foil knots the decay is described by power laws that range
form $t^{-2/3}$ to $t^{-1/3}$, which can be as slow as the $t^{-1/3}$
behavior for helical triple-ring structures that were seen in earlier work.
The two nonhelical configurations decay like $t^{-1}$, which is somewhat
slower than the previously obtained $t^{-3/2}$ behavior in the decay
of interlocked rings with zero magnetic helicity.
We attribute the difference to the creation
of local structures that contain magnetic helicity which inhibits
the field decay due to the existence of a lower bound imposed by
the realizability condition.
We show that net magnetic helicity can be produced resistively as a
result of a slight imbalance between mutually canceling helical pieces
as they are being driven apart.
We speculate that higher order topological invariants beyond magnetic
helicity may also be responsible for slowing down the decay of the two more
complicated nonhelical structures mentioned above.
\end{abstract}

\maketitle

\section{Introduction}

Magnetic helicity is an important quantity in dynamo theory
\cite{FrischPouquet1975,BrandenbSubramanianReview2005},
astrophysics \cite{RustKumar1996ApJ,Low1996SoPh}
and plasma physics
\cite{Taylor1974,Taylor1982,JensenChu1984PhFl,BergerField1984JFM}.
In the limit of high magnetic Reynolds numbers
it is a conserved quantity \cite{L.Woltjer061958}. This conservation is responsible
for an inverse cascade which can be the
cause for large-scale magnetic fields as we observe them in astrophysical
objects.
The small-scale component of magnetic helicity is responsible for the
quenching of dynamo action \cite{Gruzinov1994}
and has to be shed in order to obtain magnetic
fields of equipartition strength and sizes larger then the underlying turbulent
eddies \cite{BlackmanBrandenburg2003ApJ}.

Helical magnetic fields are observed on the Sun's
surface \cite{Seehafer1990,Pevtsov1995ApJ}.
Such fields are also produced in tokamak
experiments for nuclear fusion to contain the plasma \cite{Nelson1995PhPl}.
It could be shown that the helical
structures on the Sun's surface are more likely to erupt in coronal
mass ejections \cite{Canfield1999}, which could imply that the Sun sheds
magnetic helicity \cite{Zhang05}.
In \cite{Zhang06} it was shown that, for a force-free magnetic field
configuration, there exists an upper limit of the magnetic helicity below
which the system is in equilibrium.
Exceeding this limit leads to coronal mass ejections which drag magnetic
helicity from the Sun.

Magnetic helicity is connected with the linking of magnetic field lines.
For two separate
magnetic flux rings with magnetic flux $\phi_{1}$ and $\phi_{2}$
it can be shown that magnetic helicity is equal to twice the number of
mutual linking $n$ times the product of the two fluxes
\cite{MoffattKnottedness1969}:
\EQ \label{eq: helicity linking}
\Hm = \int_{V} \AAA\cdot\BB d V = 2n\phi_{1}\phi_{2},
\EN
where $\BB$ is the magnetic flux density, expressed in terms of the
magnetic vector potential $\AAA$ via $\BB = \nab\times\AAA$
and the integral is taken over the whole volume.
As we emphasize in this paper, however, that this formula does not
apply to the case of a single interlocked flux tube.

The presence of magnetic helicity constrains the decay of magnetic
energy \cite{L.Woltjer061958,Taylor1974} due to the
the realizability condition \cite{MoffattBook1978} which imposes a lower
bound on the spectral magnetic energy if magnetic helicity is finite; that is,
\EQ \label{eq: realizability}
M(k) \ge k|H(k)|/2\mu_{0},
\EN
where $M(k)$ and $H(k)$ are magnetic energy and helicity at
wave number $k$ and $\mu_{0}$ is the vacuum permeability.
These spectra are normalized such that
$\int M(k) d k = \langle \BB^{2}\rangle/2\mu_{0}$ and
$\int H(k) d k = \langle \AAA\cdot\BB\rangle$,
where angular brackets denote volume averages.
Note that the energy at each scale is bound separately, which constrains
conversions from large to small scales and vice versa.
For most of our calculations we assume a periodic domain
with zero net flux.
Otherwise, in the presence of a net flux, magnetic helicity would not
be conserved \cite{SMG94,Ber97}, but it would be produced at a constant
rate by the $\alpha$ effect \cite{BM04}.

The connection with the topology of the magnetic field makes the magnetic
helicity a particularly interesting quantity for studying relaxation processes.
One could imagine that the topological structure imposes limits on how magnetic
field lines can evolve during magnetic relaxation.
To test this it has been studied whether
the field topology alone can have an effect on the decay process
or if the presence of magnetic helicity is needed \cite{fluxRings10}.
The outcome was that, even for topologically nontrivial configurations, the decay
is only effected by the magnetic helicity content.
This was, however, questioned \cite{Yeates_Topology_2010}
and a topological invariant was introduced
via field line mapping which adds another constraint even in absence of magnetic
helicity.
Further evidence for the importance of extra constraints came from
numerical simulations of braided magnetic field with zero magnetic
helicity \cite{Pontin_etal11} where, at the end of a complex cascade-like
process, the system relaxed into an approximately force-free field state
consisting of two flux tubes of oppositely signed twist.
Since the net magnetic helicity is zero, the evolution of the field
would not be governed by Taylor relaxation \cite{Taylor1974} but by
extra constraints.

A serious shortcoming of some of the earlier work is that
the nonhelical field configurations considered so far were
still too simple.
For example, in the triple ring of \cite{fluxRings10} it would
have been possible to rearrange freely one of the outer rings on top
of the other one without crossing any other field lines.
The magnetic flux of these rings would annihilate to zero, making this
configuration trivially nonhelical.
Therefore, we construct in the present paper
more complex nonhelical magnetic field configurations
and study the decay of the magnetic field in a similar fashion as in
our earlier work.
Candidates for suitable field configurations
are the IUCAA logo \footnote{IUCAA = The Inter-University Centre
for Astronomy and Astrophysics in Pune, India.}
(which is a single nonhelically interlocked flux rope that will be
referred to below as the IUCAA knot)
and the Borromean rings for which $\Hm = 0$.
The IUCAA knot is commonly named $8_{18}$ in knot theory.
Furthermore, we test if Eq.\ \eqref{eq: helicity linking} is applicable for
configurations where there are no separated flux tubes while magnetic
helicity is finite. Therefore we
investigate setups where the magnetic field has the shape of a particular knot
which we call $n$--foil knot.

\section{Model}
\subsection{Representation of $n$--foil knots}

\begin{figure}[t!]\begin{center}
\includegraphics[width=0.7\columnwidth]{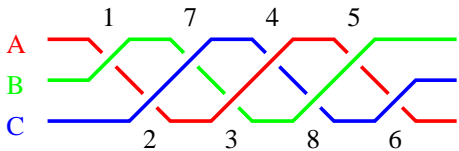}
\end{center}\caption[]{(Color online) Braid representation of the 4--foil knot.
The letters denote the starting position and the numbers denote the
crossings.}
\label{fig: 4_foil_braid}
\end{figure}

\begin{figure}[t!]\begin{center}
\includegraphics[width=0.7\columnwidth]{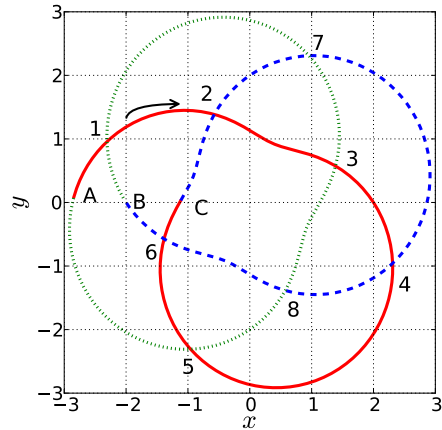}
\end{center}\caption[]{(Color online) $xy$ projection of the 4--foil knot. The
numbers denote the crossings while the colors (line styles)
separate different parts of the curve.
The letters denote the different starting positions
for the braid representation in Fig.\ \ref{fig: 4_foil_braid}. The
arrow shows the sense of the knot.}
\label{fig: 4_foil_plane_color}
\end{figure}

In topology a knot or link can be described via the braid notation
\cite{ArtinBraids1947}, where the crossings are plotted sequentially,
which results in a diagram that resembles a braid.
Some convenient starting points have
to be chosen from where the lines are drawn in the direction
according to the sense of the knot (Figs.\ \ref{fig: 4_foil_braid}
and \ref{fig: 4_foil_plane_color})
For each crossing either a capital or small letter is assigned
depending on whether it is a positive or negative crossing.

For the trefoil knot the braid representation is simply AAA. For each new
foil a new starting point is needed; at the same time
the number of crossings for each line increases by one. This means
that, for the 4--foil knot, the braid representation is ABABABAB, for the
5--foil ABCABCABCABCABC, etc.

\begin{figure}[t!]\begin{center}
\includegraphics[width=0.7\columnwidth]{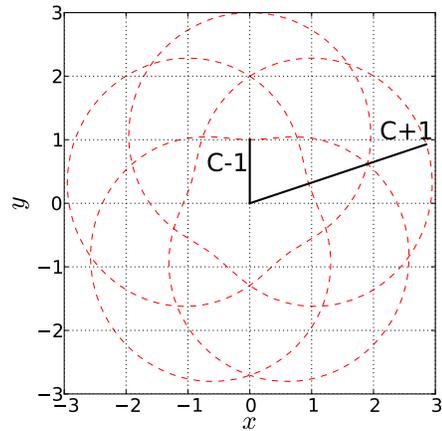}
\end{center}\caption[]{(Color online)
Projection of the 5--foil on the $xy$ plane. The lines
show the meaning of the distance $C$,
which has to be larger than $1$ to make sense.}
\label{fig: 5foil_schematic}
\end{figure}

We construct an initial magnetic field configuration in the form of an $n$--foil
knot with $\nf$ foils or leaves. First, we construct its spine or backbone as
a parametrized curve in three-dimensional space.
In analogy to \cite{PP_IdealTrefoil_01} we apply the convenient parametrization
\begin{equation}
\xx(s) = \left( \begin{array}{c}
(C+\sin{s\nf})\sin[{s(\nf-1)}] \\
(C+\sin{s\nf})\cos[{s(\nf-1)}] \\
D\cos{s\nf} \\
\end{array} \right),
\end{equation}
where $(C-1)$ is some minimum distance from the origin,
$D$ is a stretch factor in the $z$ direction
and $s$ is the curve parameter (see Fig.\ \ref{fig: 5foil_schematic}).

\begin{figure}[t!]\begin{center}
\includegraphics[width=0.5\columnwidth]{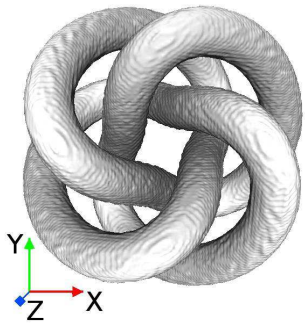}
\end{center}\caption[]{(Color online)
Isosurface of the initial magnetic field energy
for the 4--foil configuration.}
\label{fig: 4foil iso initial}
\end{figure}

The strength of the magnetic field across the tube's cross section
is constant and equal to $B_0$.
In the following we shall use $B_0$ as the unit of the magnetic field.
Since we do not want the knot to touch
itself we set $C = 1.6$ and $D = 2$.
The full three-dimensional magnetic field is constructed radially around
this curve (Fig.\ \ref{fig: 4foil iso initial}), where the thickness of the
cross section is set to $0.48$.

\subsection{The IUCAA knot}

\begin{figure}[t!]\begin{center}
\includegraphics[width=0.45\columnwidth]{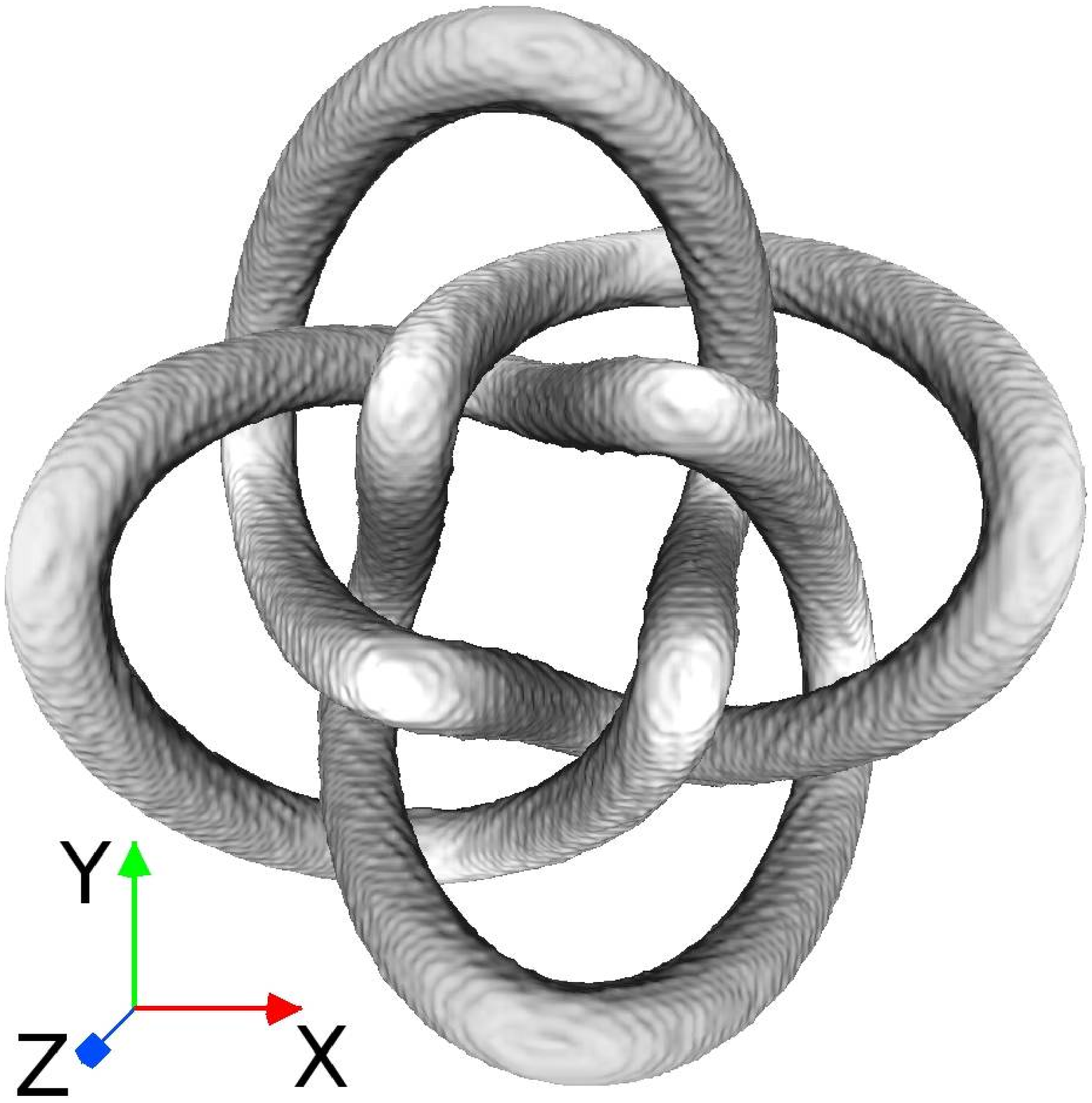}
\includegraphics[width=0.45\columnwidth]{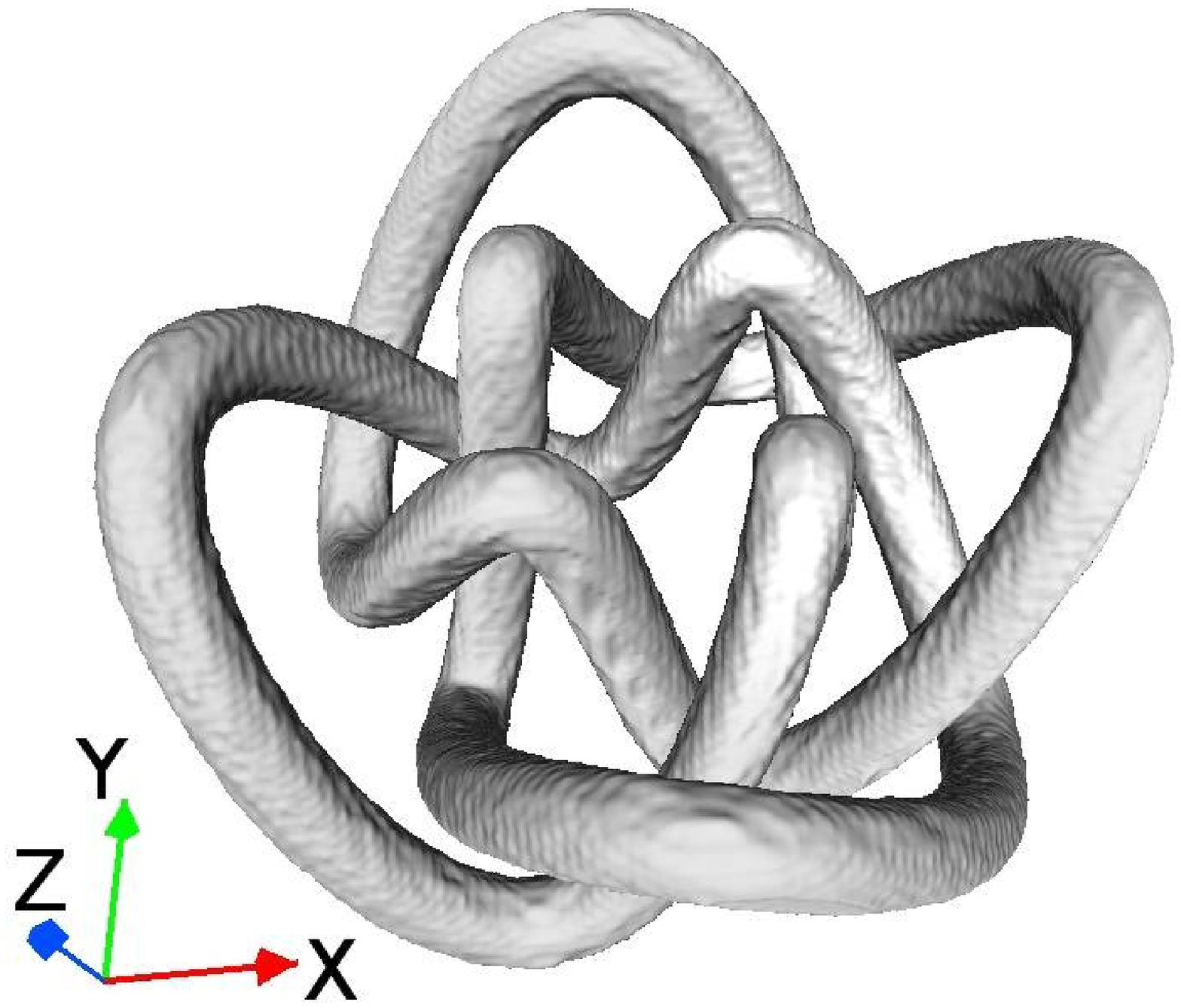}
\end{center}\caption[]{(Color online)
Isosurface of the initial magnetic field energy
for the IUCAA knot seen from the top (left panel) and slightly from the side
(right panel).}
\label{fig: iucaa_iso}
\end{figure}

A prominent example of a single nonhelically interlocked flux rope
is the IUCAA knot.
For the IUCAA knot we apply a very similar parametrization as for the $n$--foil
knots. We have to consider the faster variation in $z$ direction, which yields
\begin{equation}
\label{eq: param iucaa}
\xx(s) = \left( \begin{array}{c}
(C+\sin{4s})\sin{3s} \\
(C+\sin{4s})\cos{3s} \\
D\cos{(8s-\varphi)} \\
\end{array} \right),
\end{equation}
where $C$ and $D$ have the same meaning as for the $n$--foil knots and
$\varphi$ is a phase shift of the $z$ variation.
The full three-dimensional magnetic field is constructed radially around
this curve (Fig.\ \ref{fig: iucaa_iso}), where the thickness of the
cross section is set to $0.48$.

\subsection{Borromean rings}

\begin{figure}[t!]\begin{center}
\includegraphics[width=0.5\columnwidth]{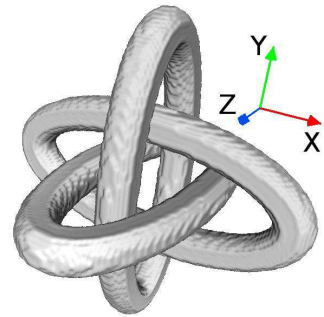}
\end{center}\caption[]{(Color online)
Isosurface of the initial magnetic field energy
for the Borromean rings configuration.}
\label{fig: borromean iso initial}
\end{figure}

The Borromean rings are constructed with three ellipses whose surface
normals point in the direction of the unit vectors
(Fig.\ \ref{fig: borromean iso initial}).
The major and minor axes are set to $2.5$ and $1$, respectively,
and the thickness of the cross section is set to $0.6$.
If any one of the three rings were removed, the remaining 2 rings
would no longer be interlocked.
This means that there is no mutual
linking and hence no magnetic helicity. One should, however, not consider this
configuration as topologically trivial, since the rings cannot be separated,
which is reflected in a nonvanishing third-order topological invariant
\cite{ruzmaikin:331}.

\subsection{Numerical setup}

We solve the resistive magnetohydrodynamical (MHD) equations
for an isothermal compressible gas,
where the gas pressure is given by $p=\rho c_{S}^{2}$, with the density
$\rho$ and isothermal sound speed $c_{S}$. Instead of solving for
the magnetic field $\BB$ we solve for its vector potential $\AAA$
and choose the resistive gauge, since it is numerically well behaved
\cite{helicityAdvective11}.
The equations we solve are
\EQ
\frac{\partial \AAA}{\partial t} = \UU \times \BB + \eta\nabla^{2}\AAA,
\label{dAdt}
\EN
\EQ
\frac{\DD \UU}{\DD t} = -c_{\rm S}^{2} \nab \ln{\rho} +
\JJ\times\BB/\rho + \FF_{\rm visc},
\label{dUdt}
\EN
\EQ
\frac{\DD \ln{\rho}}{\DD t} = -\nab \cdot \UU,
\label{drhodt}
\EN
where $\UU$ is the velocity field, $\eta$ is the magnetic diffusivity,
$\JJ = \nab\times\BB/\mu_{0}$ is the current density,
$\FF_{\rm visc} = \rho^{-1}\nab\cdot2\nu\rho\SSSS$ is the viscous
force with the traceless rate of strain tensor $\SSSS$ with components
${\sf S}_{ij}=\frac{1}{2}(u_{i,j}+u_{j,i})-\frac{1}{3}\delta_{ij}\nab\cdot\UU$,
$\nu$ is the kinematic viscosity, and
$\DD/\DD t=\partial/\partial t + \UU\cdot\nabla$ is the advective
time derivative.
We perform simulations in a box of size $(2\pi)^{3}$ with fully periodic
boundary conditions for all quantities. To test how boundary effects play a role
we also perform simulations with perfect conductor boundary conditions (i.e.,
the component of the magnetic field perpendicular to the surface vanishes).
In both choices of boundary conditions, magnetic helicity is gauge invariant
and is a conserved quantity in ideal MHD (i.e., $\eta=0$).
As a convenient parameter we use the Lundquist number
$\Lu = U_{\rm A}L/\eta$, where $U_{\rm A}$ is the Alfv\'en velocity
and $L$ is a typical length scale of the system.
The value of the viscosity is characterized by the magnetic Prandtl
number $\Pm=\nu/\eta$.
However, in all cases discussed below we use $\Pm=1$.
To facilitate comparison of different setups it is convenient to
normalize time by
the resistive time $t_{\rm res} = r^{2}\pi/\eta$, where $r$ is the
radius of the cross section of the flux tube.

We solve \Eqss{dAdt}{drhodt} with the {\textsc Pencil Code} \cite{BD02PC},
which employs
sixth-order finite differences in space and a third-order time
stepping scheme.
As in our earlier work \cite{fluxRings10}, we use $256^3$ meshpoints
for all our calculations.
We recall that we use explicit viscosity and magnetic diffusivity.
Their values are dominant over numerical contributions associated
with discretization errors of the scheme \footnote{The
discretization error of the temporal scheme scheme implies a
small diffusive contribution proportional to $\nabla^4$, but even
at the Nyquist frequency this is subdominant.}.

\section{RESULTS}

\subsection{Helicity of $n$--foil knots}

\begin{figure}[t!]\begin{center}
\includegraphics[width=\columnwidth]{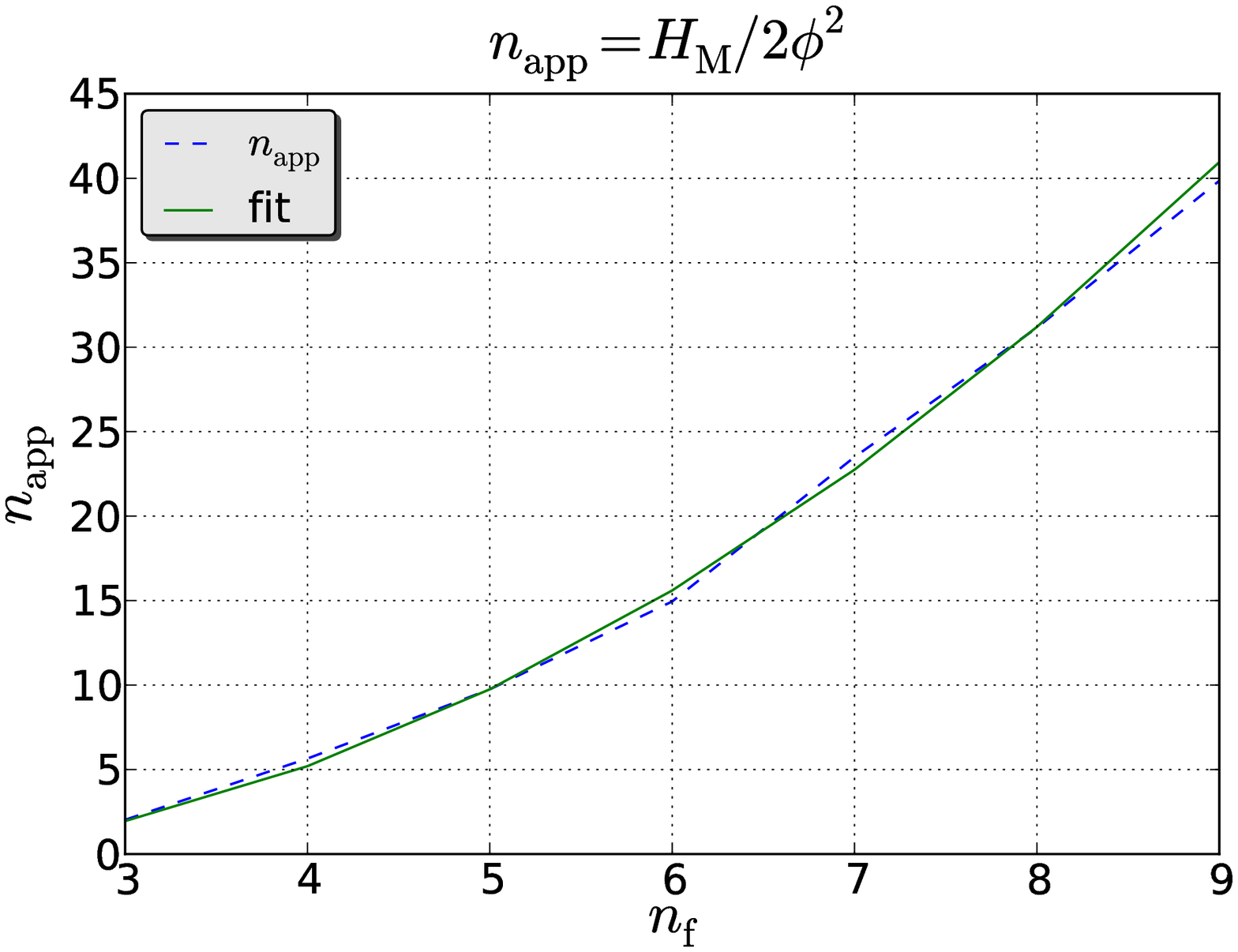}
\includegraphics[width=\columnwidth]{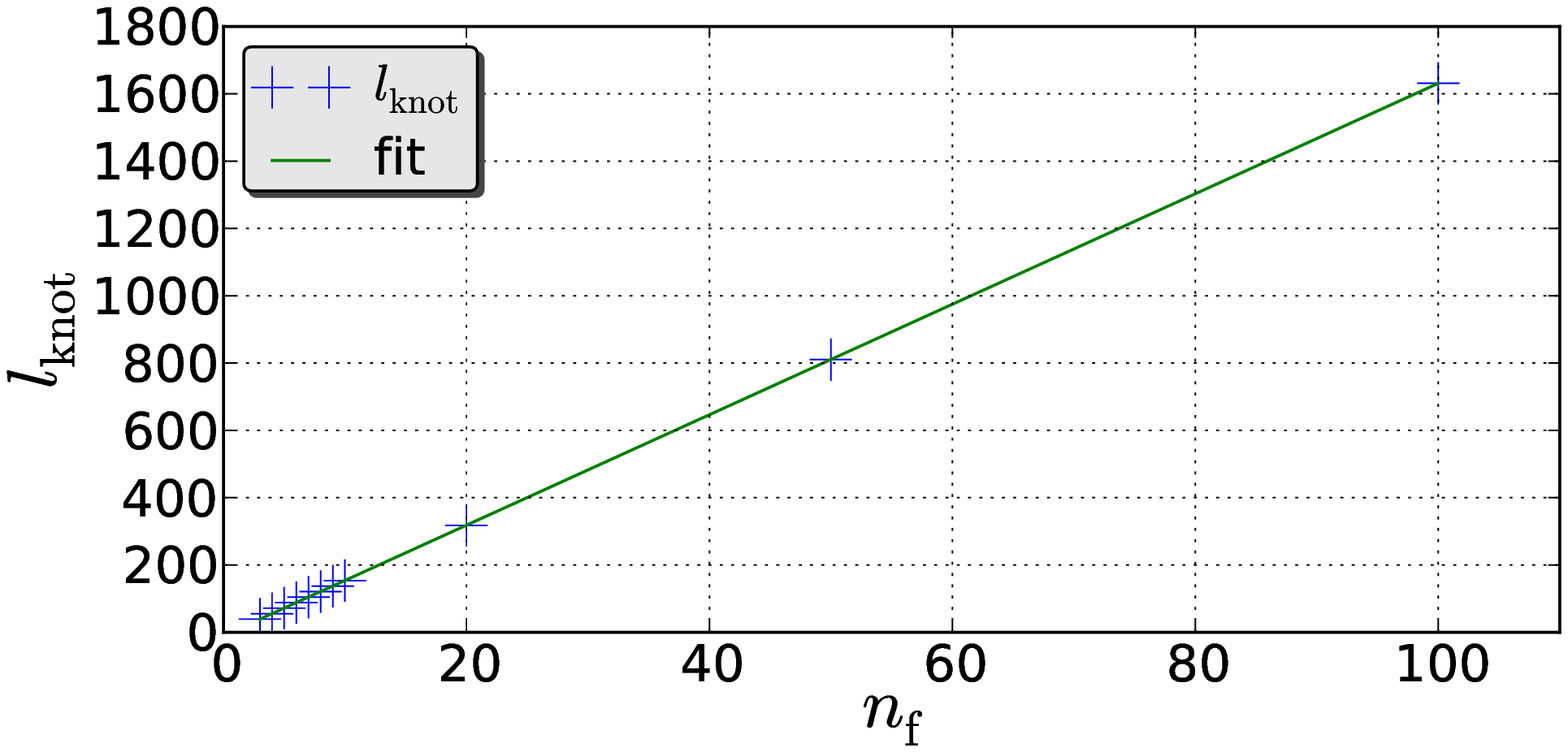}
\end{center}\caption[]{(Color online) The apparent self-linking number
for $n$--foil knots with respect to $\nf$ (upper panel).
The fit is obtained by equating
\eqref{eq: helicity linking} and \eqref{eq: H_nf}.
The length of a $n$--foil knot is plotted with respect to $\nf$
(lower panel), which can be fit almost perfectly by a linear function.
}
\label{fig: n_apparent}
\end{figure}

We test equation \eqref{eq: helicity linking} for the $n$--foil knots in order
to see how the number of foils $\nf$ relates to the number of mutual linking
$n$ for the separated flux tubes.
From our simulations we know the
magnetic helicity $\Hm$ and the magnetic flux $\phi$ through the tube.
Solving \eqref{eq: helicity linking} for $n$ will lead to an apparent
self-linking number which we call $\napp$. It turns out that
$\napp$ is much larger then $\nf$ and increases faster
(Fig.\ \ref{fig: n_apparent}).

We note that \eqref{eq: helicity linking} does not apply to this setup
of flux tubes and propose therefore a different formula for the magnetic
helicity:
\begin{eqnarray} \label{eq: H_nf}
\Hm & = & (\nf-2)\nf \phi^{2}/2.
\end{eqnarray}
In \Fig{fig: n_apparent} we plot the apparent linking number together
with a fit which uses equation \eqref{eq: H_nf}.

\begin{figure}[t!]\begin{center}
\includegraphics[width=\columnwidth]{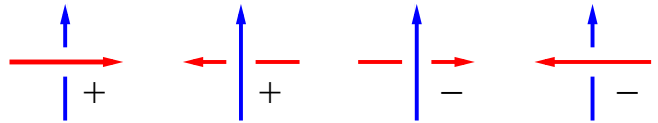}
\end{center}\caption[]{(Color online) Schematic representation
illustrating the sign of a crossing.
Each crossing has a handedness which can be either positive or negative. The
sum of the crossings gives the number of linking and eventually the magnetic
helicity content via equation \eqref{eq: H_nf}.}
\label{fig: crossings schematically}
\end{figure}

\begin{figure}[t!]\begin{center}
\includegraphics[width=0.7\columnwidth]{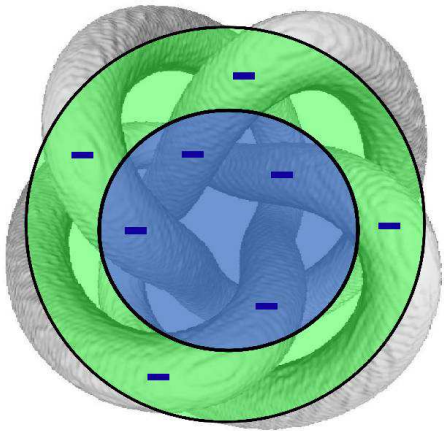}
\end{center}\caption[]{(Color online) The isosurface for the 4--foil knot field
configuration. The sign of the crossing is always negative. The
rings show the different areas where crossings occur.}
\label{fig: 4foil_crossings}
\end{figure}

Equation \eqref{eq: H_nf} can be motivated via the number of crossings.
The flux tube is projected onto the $xy$ plane such that the number of crossings
is minimal. The linking number can be determined by adding all
positive crossings and subtracting all negative crossings according
to \Fig{fig: crossings schematically}.
The linking number is then simply given as \cite{SchareinPhD}
\EQ
n_{\rm linking} = (n_{+}-n_{-})/2,
\EN
where $n_{+}$ and $n_{-}$ correspond to positive and negative crossings,
respectively. If we set $n_{\rm linking} = n_{\rm app}$ then we
easily see the validation of \eqref{eq: H_nf}.
Each new foil creates a new ring of crossings and adds up one crossing
in each ring (see Fig. \ref{fig: 4foil_crossings}), which explains the
quadratic increase.

\subsection{Magnetic energy decay for $n$--foil knots}

\begin{figure}[t!]\begin{center}
\includegraphics[width=\columnwidth]{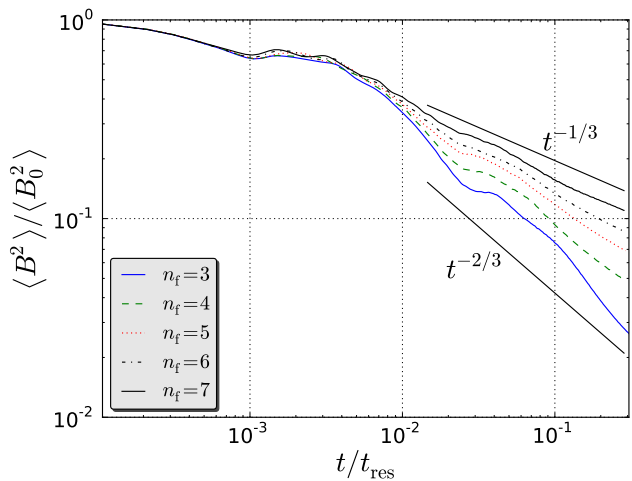}
\end{center}\caption[]{(Color online) Time dependence of the normalized
magnetic energy for a given number of foils with
periodic boundary conditions.
The power law for the energy decay varies between $-2/3$
for $\nf=3$ (solid/blue line) and $-1/3$ for $\nf=7$
(solid/black).}
\label{fig: magnetic energy nfoil}
\end{figure}

Next, we plot in \Fig{fig: magnetic energy nfoil}
the magnetic energy decay for $n$--foil knots with $\nf=3$ up to
$\nf=7$ for periodic boundary conditions.
It turns out that, at later times, the decay slows down as $\nf$ increases.
The decay of the magnetic energy obeys an approximate $t^{-2/3}$ law
for $\nf=3$ and a $t^{-1/3}$ law for $\nf=7$.
The rather slow decay is surprising in view of earlier results that,
for turbulent magnetic fields, the magnetic energy decays like $t^{-1}$
in the absence of magnetic helicity and like $t^{-1/2}$ with magnetic
helicity \cite{CHB05}.
Whether or not the decay seen in \Fig{fig: magnetic energy nfoil} really
does follow a power law with such an exponent remains therefore open.

\begin{figure}[t!]\begin{center}
\includegraphics[width=\columnwidth]{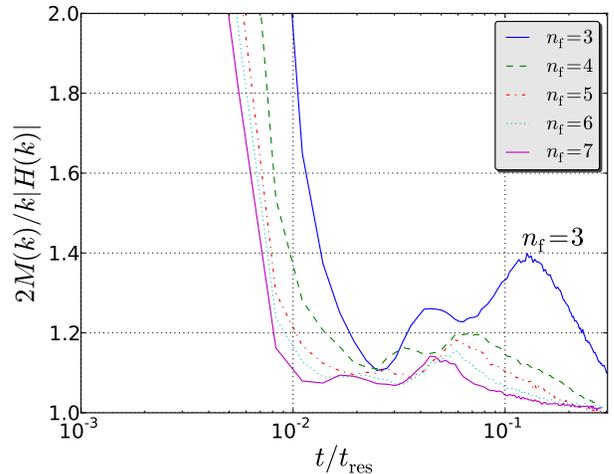}
\end{center}\caption[]{
(Color online) Time dependence of the quotient from the realizability
condition \eqref{eq: realizability} for $k=2$.
It is clear that, for larger $\nf$, the energy approaches its minimum faster.
}
\label{fig: realizability_k2}
\end{figure}

\begin{figure}[t!]\begin{center}
\includegraphics[width=\columnwidth]{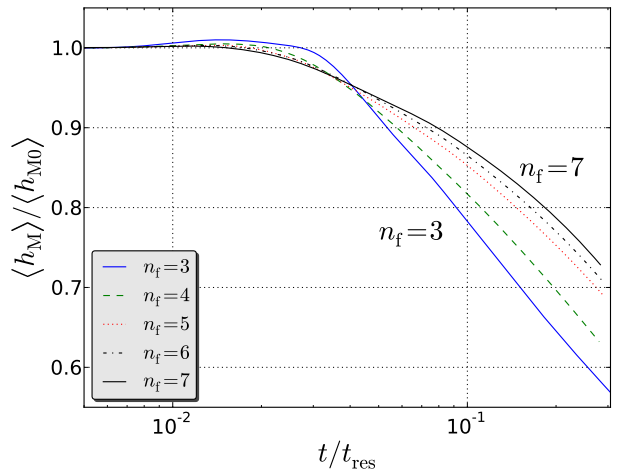}
\end{center}\caption[]{(Color online) Time dependence of the normalized
magnetic helicity for a given number of foils with
periodic boundary conditions.}
\label{fig: mag hel decay}
\end{figure}

The different power laws for a given number of foils $\nf$ are unexpected
because the setups differ only in their magnetic helicity and
magnetic energy content and not in the qualitative nature of the knot.
Indeed, one might have speculated that the faster $t^{-2/3}$ decay applies
to the case with larger $\nf$, because this structure is more complex
and involves sharper gradients.
On the other hand, a larger value of $\nf$ increases the total helicity,
making the resulting knot more strongly packed.
This can be verified by noting that the magnetic
helicity increases quadratically with $\nf$ while the magnetic energy
increases only linearly.
This is because the energy is proportional to the length
of the tube which, in turn, is proportional to $\nf$ (\Fig{fig: n_apparent}).
Therefore we expect that, for the higher $\nf$ cases,
the realizability condition should play a more significant role
at early times.
This can be seen in \Fig{fig: realizability_k2},
where we plot the ratio $2M(k)/(k|H(k)|)$
for $\nf = 3$ to $\nf = 7$ for $k = 2$. Since the magnetic helicity
relative to the magnetic energy is higher for larger values of $\nf$, it plays
a more significant role for high $\nf$.
This would explain a different onset of the power law
decay, although it would not explain a change in the exponent.
Indeed the decay of $\Hm$ shows approximately the same behavior for
all $\nf$ (\Fig{fig: mag hel decay}).
We must therefore expect that the different decay laws
are described only approximately by power laws.

\begin{figure}[t!]\begin{center}
\includegraphics[width=\columnwidth]{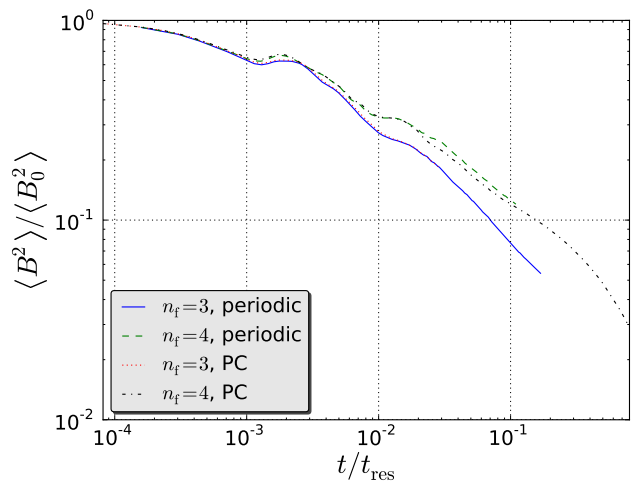}
\end{center}\caption[]{(Color online) Time dependence of the normalized
magnetic energy
for the trefoil and 4--foil knot with periodic and perfect
conductor (PC) boundary conditions. There is no significant difference
in the energy decay for the different boundary conditions.}
\label{fig: magnetic energy nfoil PC}
\end{figure}

For periodic boundary conditions it is possible that the flux tube reconnects
over the domain boundaries which could lead to additional magnetic field
destruction.
To exclude such complications we compare simulations with perfectly
conducting or closed boundaries with periodic boundary conditions
(Fig.\ref{fig: magnetic energy nfoil PC}).
Since there is no difference in the two cases we can exclude the 
significance of boundary effects for the magnetic energy decay.

\begin{figure}[t!]\begin{center}
\includegraphics[width=0.8\columnwidth]{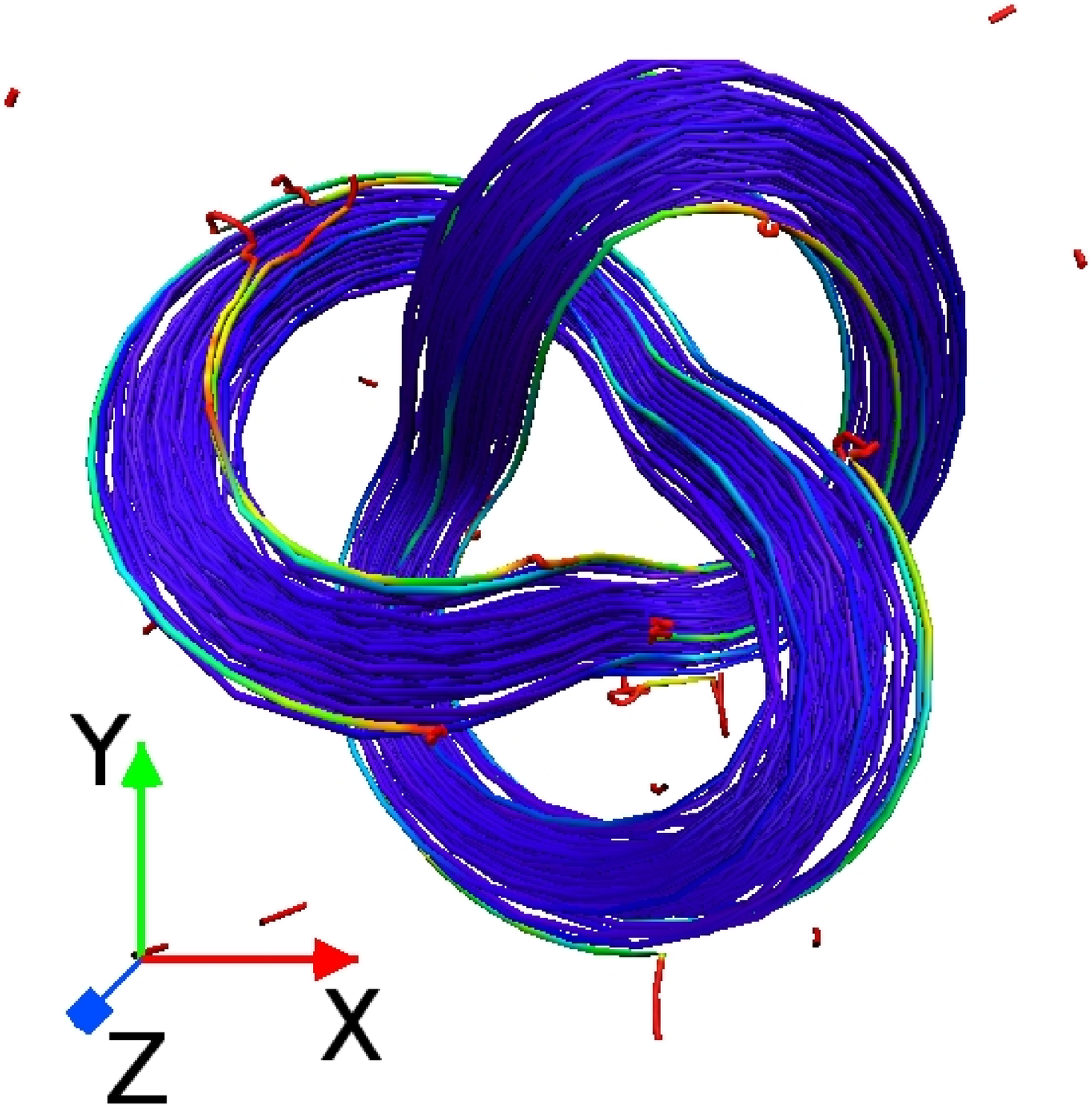}
\includegraphics[width=0.8\columnwidth]{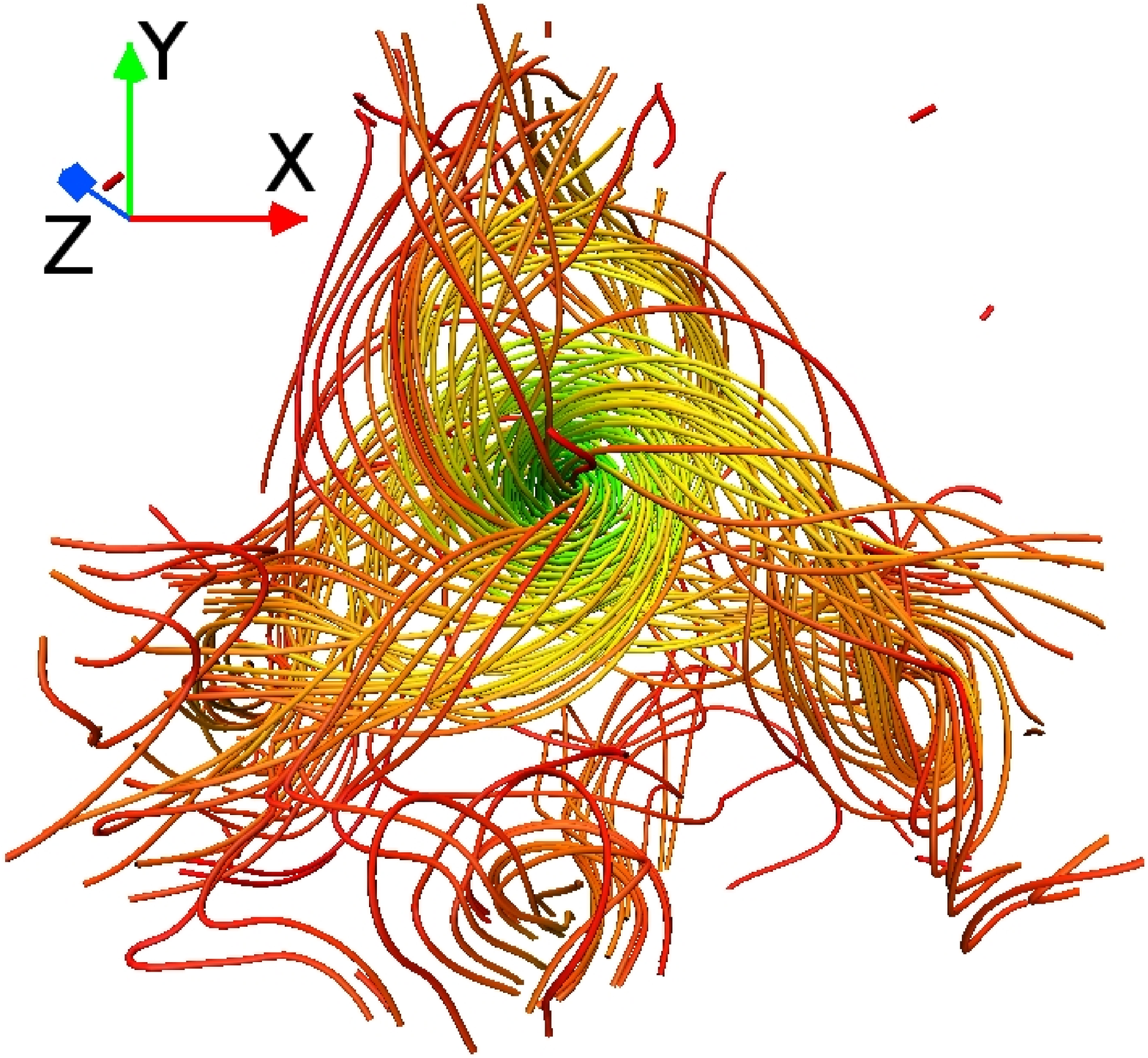}
\end{center}\caption[]{(Color online) Magnetic field lines for the trefoil knot
at time $t = 0$ (upper panel) and
$t=7.76\times10^{-2}\,t_{\rm res}$ (lower panel).
Both images were taken from the same viewing position to make
comparisons easier. The Lundquist number was chosen to be $1000$.
The colors indicate the field strength. 
}
\label{fig: trefoil0}
\end{figure}

In all cases the magnetic helicity can only decay on a resistive time
scale (Fig.\ \ref{fig: mag hel decay}). This means that, during
faster dynamical processes like magnetic reconnection, magnetic helicity
is approximately conserved. To show this we plot the magnetic field lines
for the trefoil knot at different times
(Fig.\ \ref{fig: trefoil0}). Since magnetic helicity
does not change significantly, the self-linking is transformed into
a twisting of the flux tube which is topologically equivalent to linking.
Such a process has also been mentioned in connection with Fig.~1 of
Ref.~\cite{PhysRevLett.83.1155}, while the opposite process of the conversion
of twist into linkage has been seen in Ref.~\cite{Linton_etal01}.
We can also see that the reconnection process, which transforms the trefoil
knot into a twisted ring, does not aid the decay of magnetic helicity.
    
\subsection{Decay of the IUCAA knot}

\begin{figure}[t!]\begin{center}
\includegraphics[width=0.9\columnwidth]{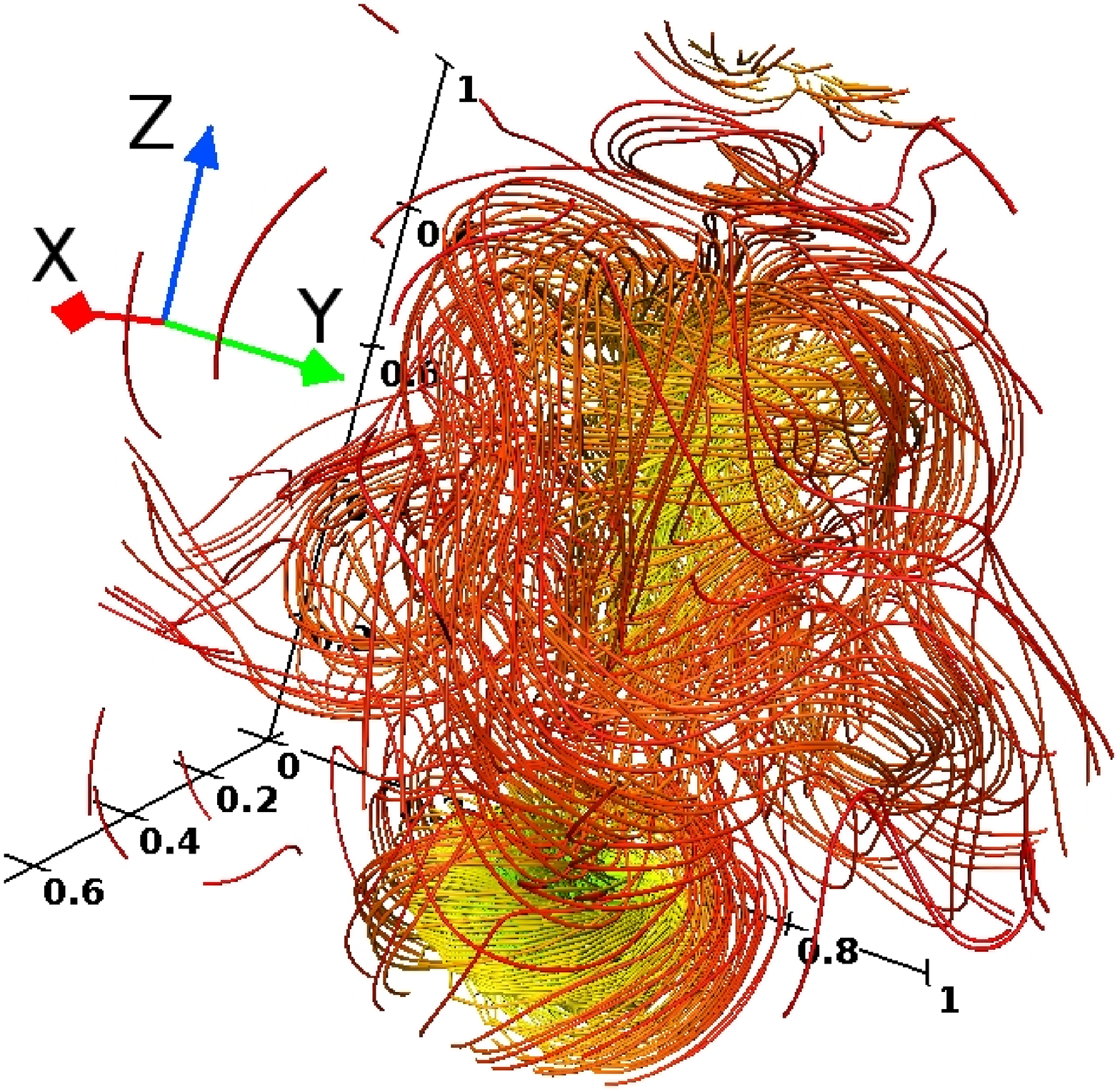}
\includegraphics[width=0.9\columnwidth]{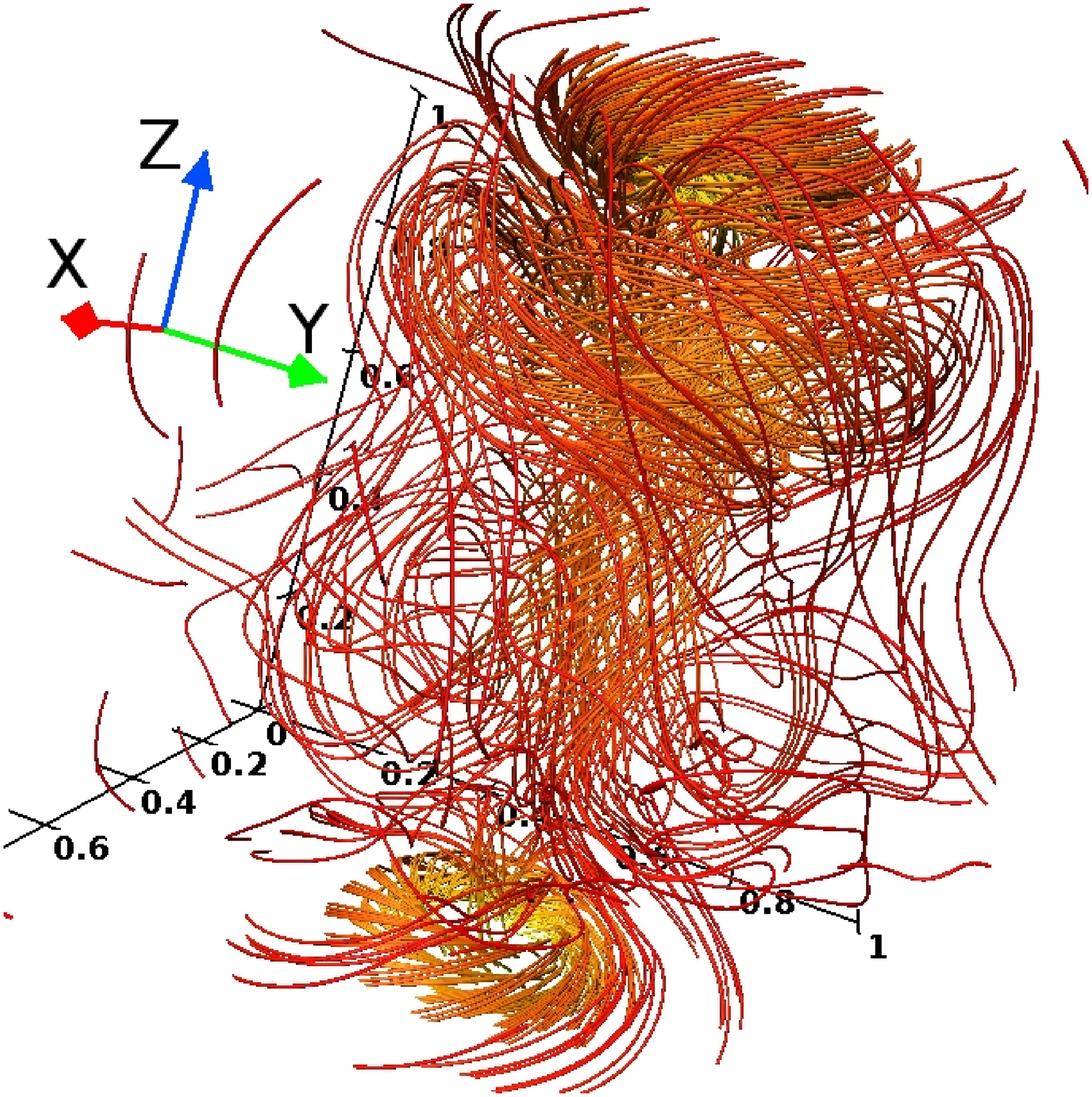}
\end{center}\caption[]{(Color online) Magnetic field lines for the IUCAA knot at
$t=0.108\,t_{\rm res}$
(upper panel) and at $t=0.216\,t_{\rm res}$ (lower panel)
for $\Lu = 1000$ and $\varphi = (4/3)\pi$.}
\label{fig: iucaa_256a1_t39}
\end{figure}

For the nonhelical triple-ring configuration of Ref.~\cite{fluxRings10}
it was found that
the topological structure gets destroyed after only
$10$ Alfv\'en times.
The destruction was attributed to the absence of magnetic
helicity whose conservation would pose constraints on the relaxation process.
Looking at the magnetic field lines of the IUCAA knot at different times
(\Fig{fig: iucaa_256a1_t39}), we see that the field remains structured
and that some helical features emerge above and below the $z=0$ plane.
These localized helical patches could then locally impose constraints
on the magnetic field decay.

\begin{figure}[t!]\begin{center}
\includegraphics[width=\columnwidth]{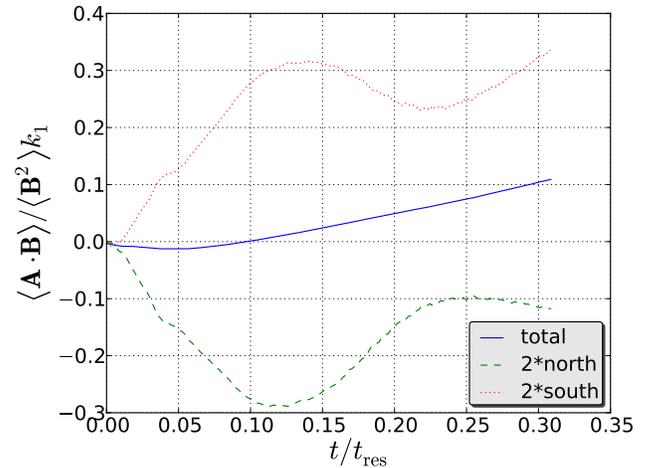}
\end{center}\caption[]{(Color online)
Normalized magnetic helicity in the northern (green/dashed line) and southern
(red/dotted line) domain half together with the total magnetic helicity
(blue/solid line) for the IUCAA knot with $\Lu = 2000$ and $\varphi = (4/3)\pi$.}
\label{fig: helicity iucaa 256c}
\end{figure}

The asymmetry of the IUCAA knot in the $z$ direction leads to different
signs of magnetic helicity above and below the $z=0$ plane.
This is shown in \Figs{fig: helicity iucaa 256c}{fig: helicity iucaa 256c2}
where we plot the magnetic helicity for the upper and lower
parts for two different values of $\varphi$; see \Eq{eq: param iucaa}.
In the plot, we refer to the upper and lower parts as north and south,
respectively.
These plots show that
there is a tendency of magnetic helicity of opposite sign to emerge
above and below the $z=0$ plane.
Given that the magnetic helicity was initially zero, one
may speculate that higher order topological invariants could
provide an appropriate tool to characterize the emergence of
such a ``bi-helical'' structure from an initially nonhelical one.

\begin{figure}[t!]\begin{center}
\includegraphics[width=\columnwidth]{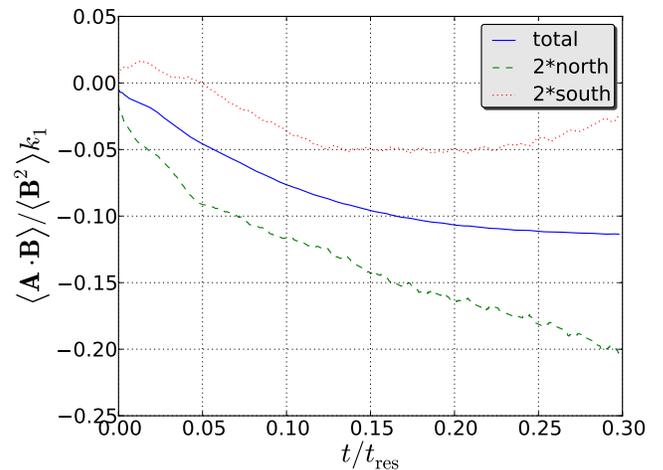}
\end{center}\caption[]{(Color online)
Normalized magnetic helicity in the northern (green/dashed line) and southern
(red/dotted line) domain half together with the total magnetic helicity
(blue/solid line) for the IUCAA knot with $\Lu = 2000$ and $\varphi = (4/3+0.2)\pi$.}
\label{fig: helicity iucaa 256c2}
\end{figure}

Note that there is a net increase of magnetic helicity over the
full volume.
Furthermore, the initial magnetic helicity is not exactly zero either,
but this is probably a consequence of discretization errors
associated with the initialization.
The subsequent increase of magnetic helicity can only occur on the longer
resistive time scales, since magnetic helicity is conserved on dynamical
time scales.
Note, however, that the increase of magnetic helicity is exaggerated
because we divide by the mean magnetic energy density which is decreasing
with time.

\begin{figure}[t!]\begin{center}
\includegraphics[width=\columnwidth]{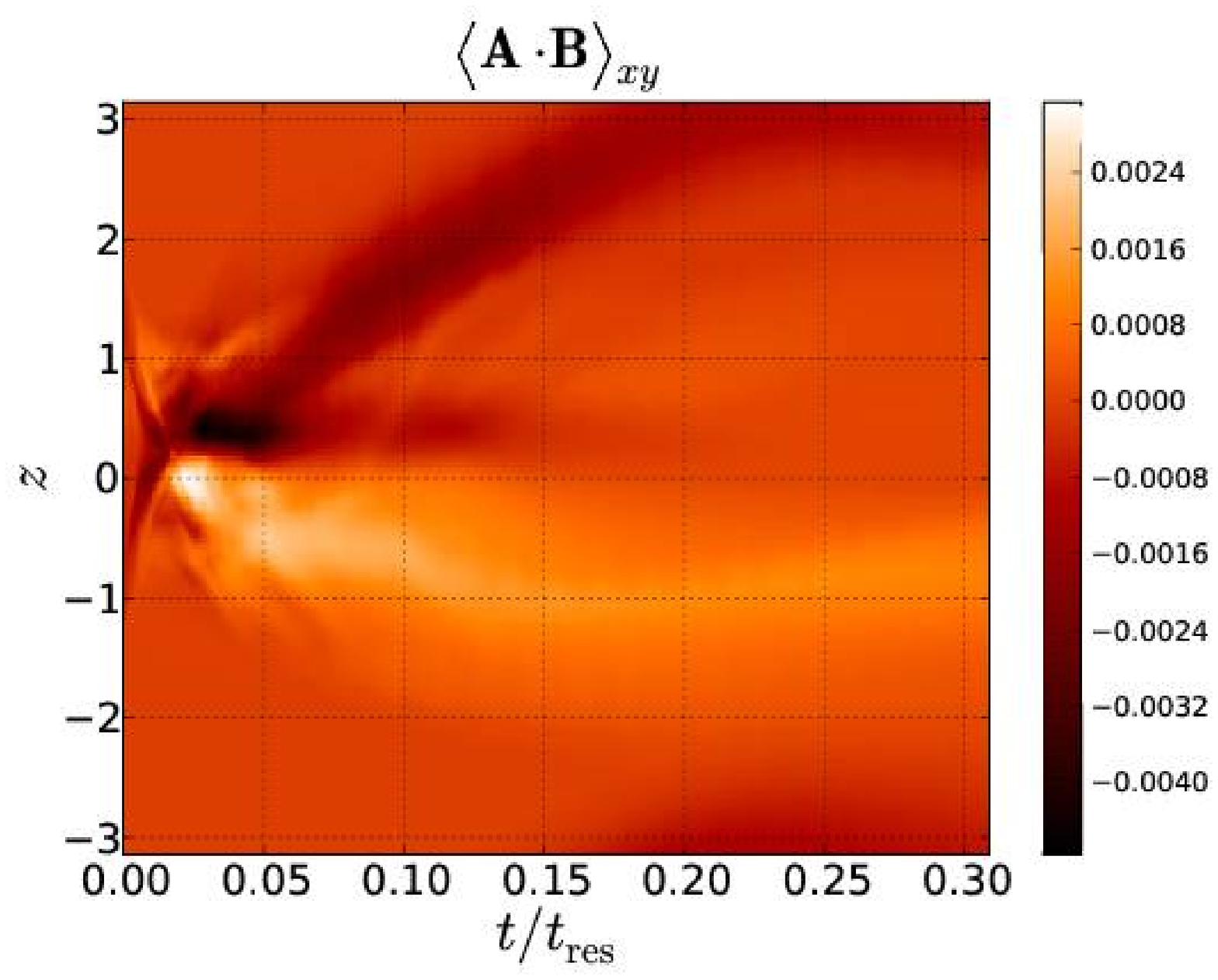}
\end{center}\caption[]{(Color online)
$xy$-averaged magnetic helicity density profile in
$z$ direction for the IUCAA knot with $\Lu = 2000$ and $\varphi = (4/3)\pi$.
There is an apparent asymmetry in the distribution amongst the hemispheres.}
\label{fig: helicity iucaa xyaveraged 256c2}
\end{figure}

In \Fig{fig: helicity iucaa xyaveraged 256c2} we plot the $xy$-averaged
magnetic helicity as a function of $z$ and $t$.
This shows that the asymmetry between upper and lower parts
increases with time, which we attribute
to the Lorentz forces through which the knot shrinks and compresses its interior.
This is followed by the ejection of magnetic field.

\begin{figure}[t!]\begin{center}
\includegraphics[width=\columnwidth]{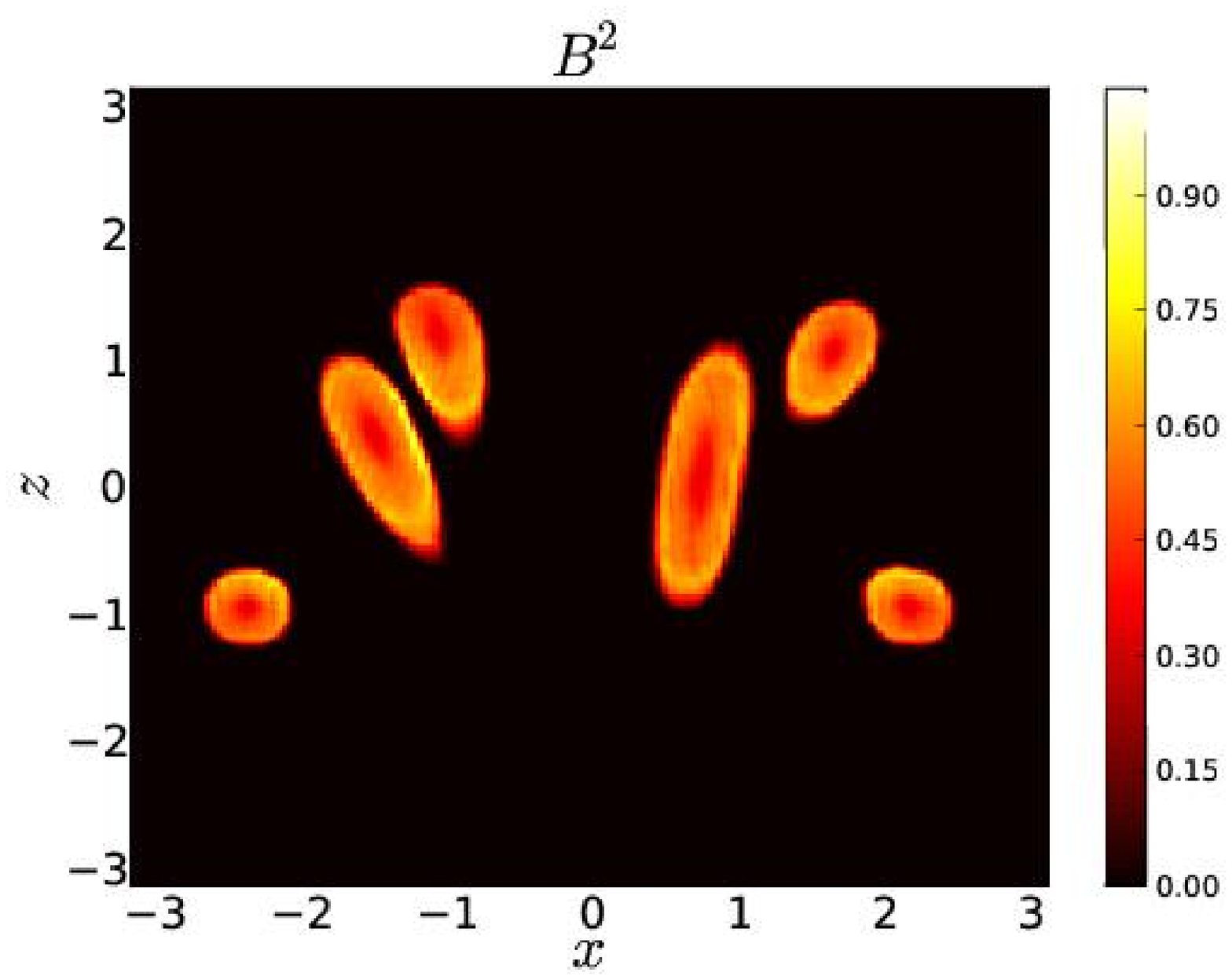} \\
\includegraphics[width=\columnwidth]{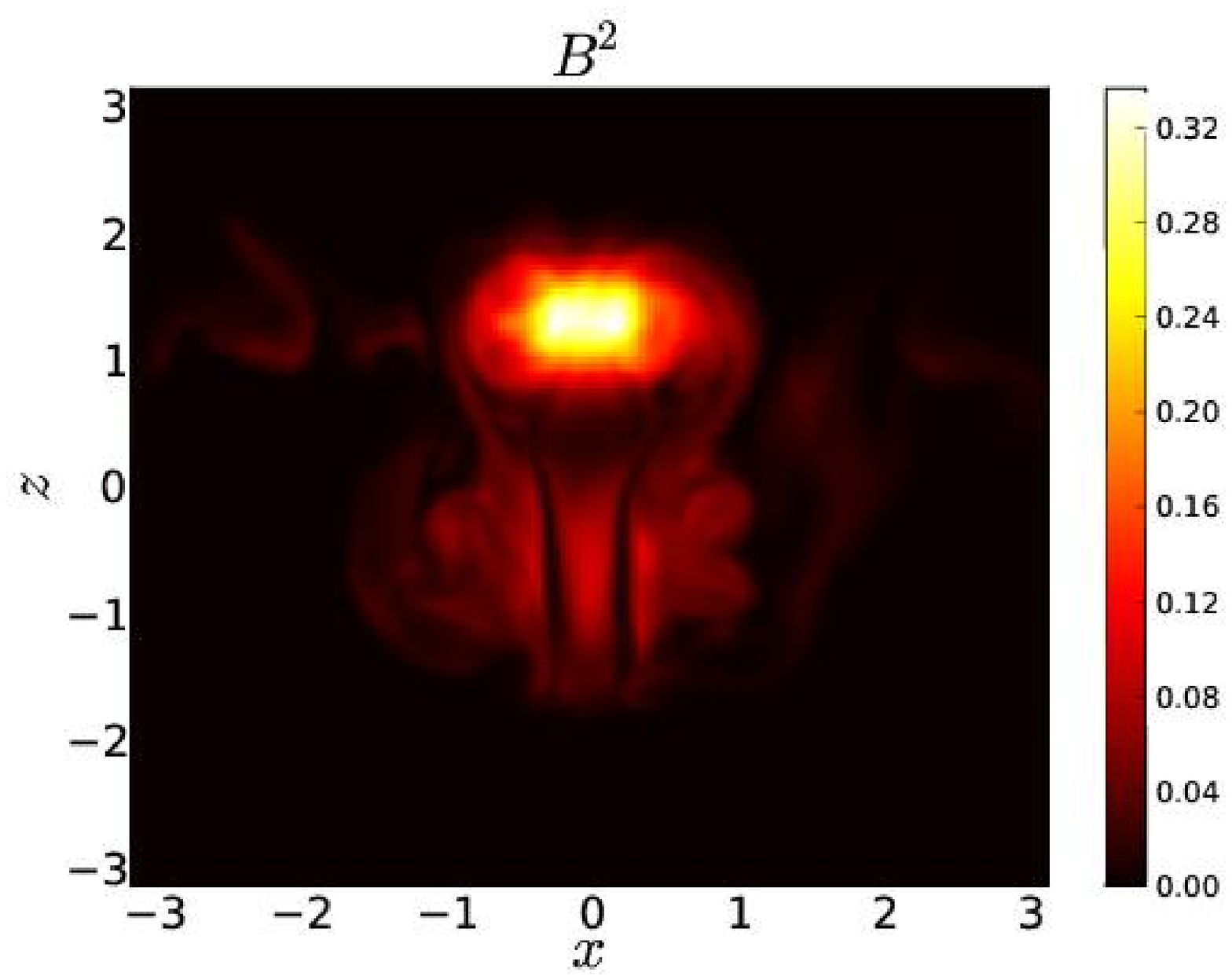}
\end{center}\caption[]{
(Color online) Magnetic energy density in the $xz$ plane for $y=0$ at $t=0$
(upper panel) and $t=5.58\times10^{-2}\,t_{\rm res}$ (lower panel) for
the IUCAA knot with
$\Lu = 2000$ and $\varphi = (4/3)\pi$.
}
\label{fig: energy_slice_c}
\end{figure}

\begin{figure}[t!]\begin{center}
\includegraphics[width=\columnwidth]{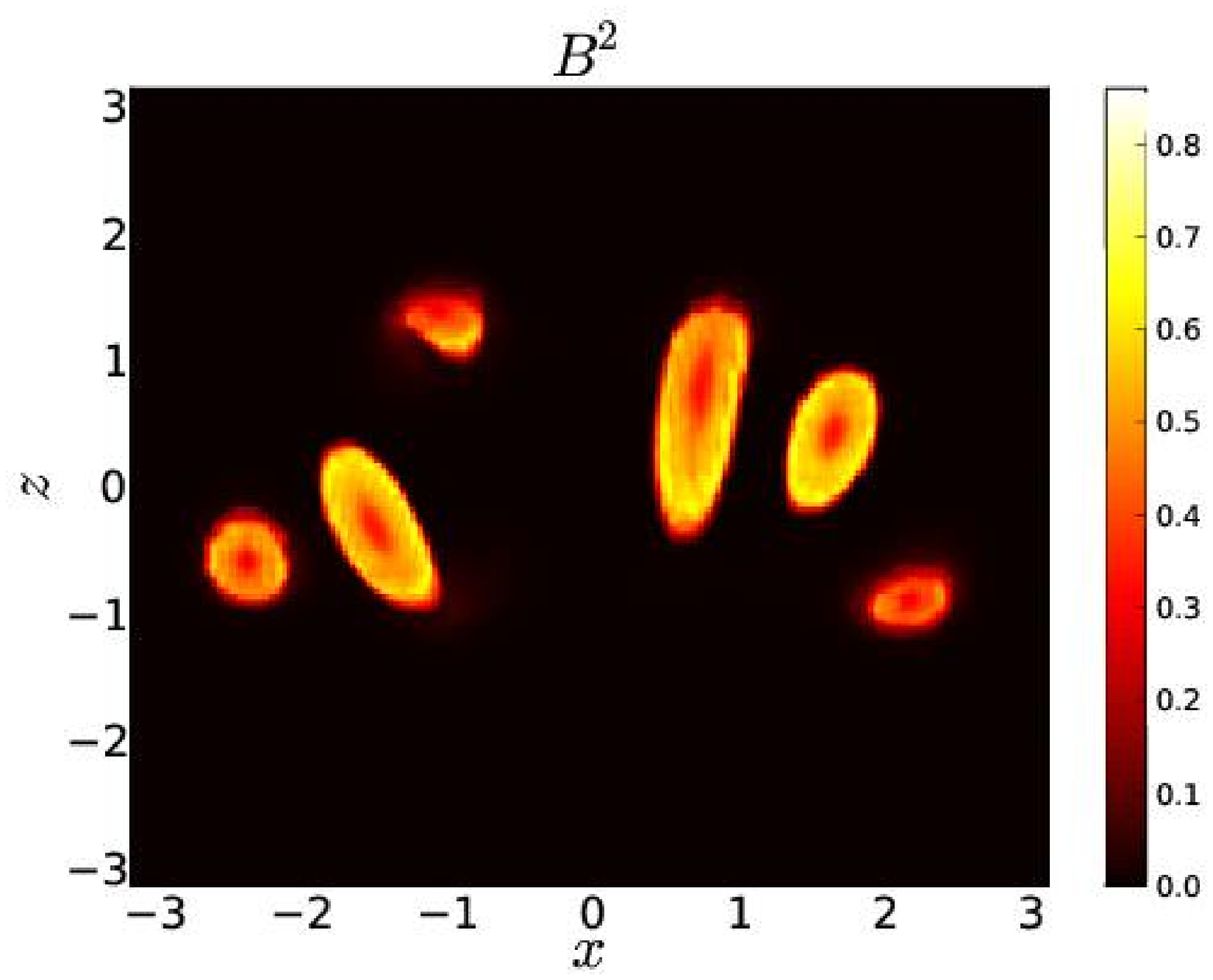} \\
\includegraphics[width=\columnwidth]{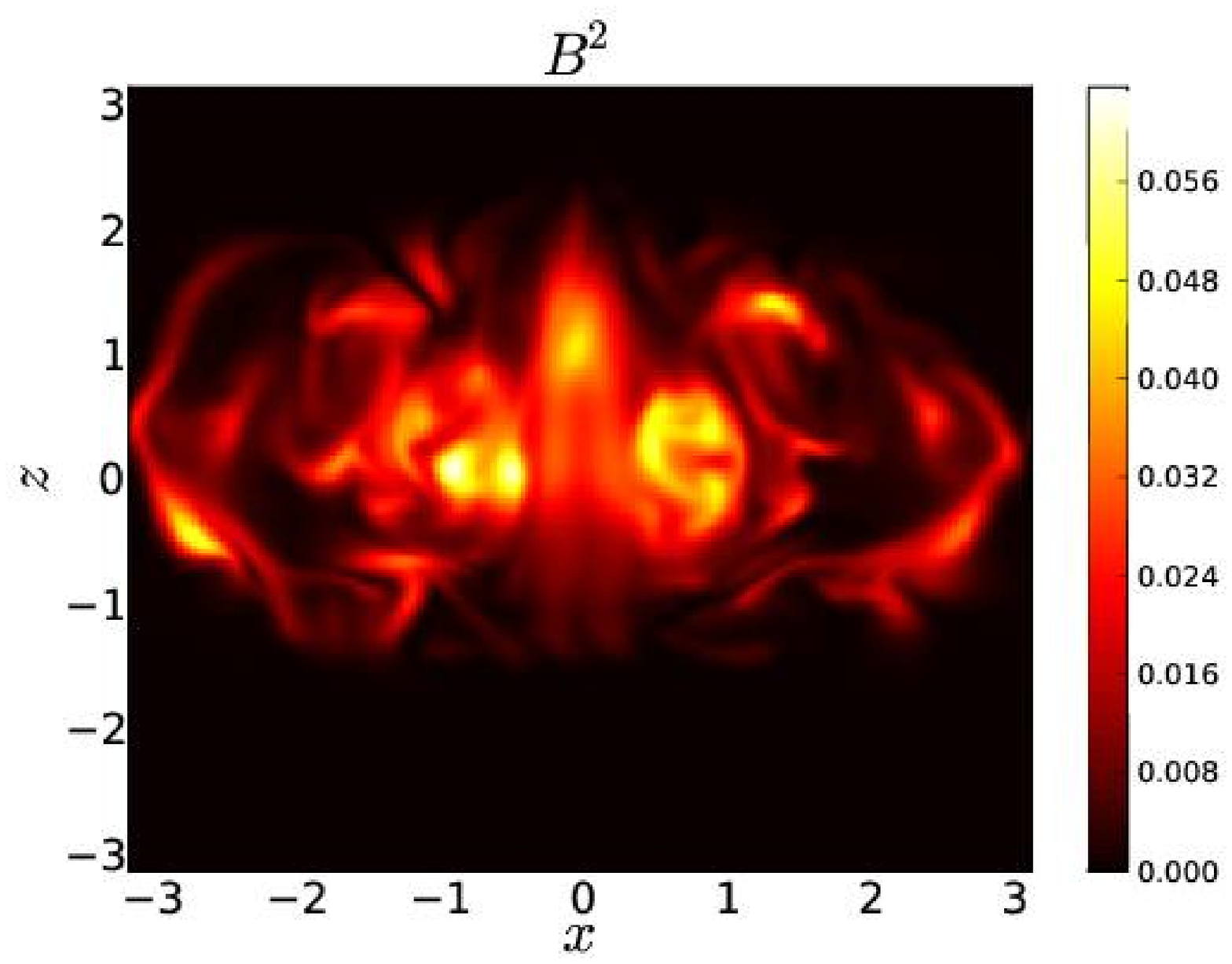}
\end{center}\caption[]{(Color online)
Magnetic energy density in the $xz$-plane for $y=0$ at $t=0$
(upper panel) and $t=5.58\times10^{-2}\,t_{\rm res}$ (lower panel)
for the IUCAA knot with
$\Lu = 2000$ and $\varphi = (4/3+0.2)\pi$. The magnetic field stays centered.}
\label{fig: energy_slice_c2}
\end{figure}

To clarify this we plot slices of the magnetic
energy density in the $xz$ plane for different
times (\Fig{fig: energy_slice_c}).
The slices are set in the center of the domain.
Due to the rose-like shape, our representation of the IUCAA knot is
not quite symmetric and turns out to be narrower in the lower half
(negative $z$) than in the upper half (positive $z$), which
is shown in \Fig{fig: iucaa_iso} (right panel).
When the knot contracts due to the Lorentz force, it begins to touch the
inner parts which creates motions in the positive $z$ direction which,
in turn, drag the magnetic field away from the center (\Fig{fig: energy_slice_c}).
The pushing of material can, however, be decreased when the phase $\varphi$
is changed.
For $\varphi = (4/3+0.2)\pi$ there is no such upward motion visible
and the configuration stays nearly symmetric (\Fig{fig: energy_slice_c2}).

In \Fig{fig: energy decay compare} the decay behavior of the magnetic energy
is compared with previous work \cite{fluxRings10}.
We note in passing that
the power law of $t^{-1}$ is expected for nonhelical turbulence \cite{CHB05},
but it is different from the
helical ($t^{-1/2}$) and nonhelical ($t^{-3/2}$) triple-ring configurations
studied earlier.
A possible explanation is the conservation of magnetic structures for
the IUCAA knot, whereas the nonhelical triple-ring configuration loses its
structure.

\subsection{Borromean rings}

\begin{figure}[t!]\begin{center}
\includegraphics[width=\columnwidth]{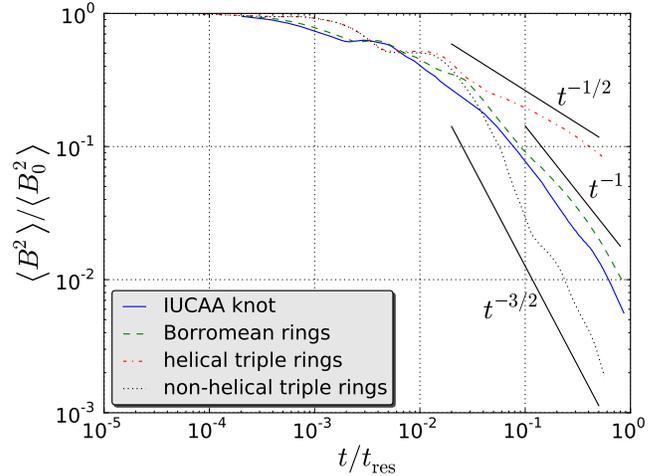}
\end{center}\caption[]{(Color online) Magnetic energy versus time for the
different initial field configurations together with power
laws which serve as a guide.
The decay speed of the IUCAA knot and Borromean rings lies
well in between the helical and nonhelical triple-ring configuration.
}
\label{fig: energy decay compare}
\end{figure}

\begin{figure}[t!]\begin{center}
\includegraphics[width=\columnwidth]{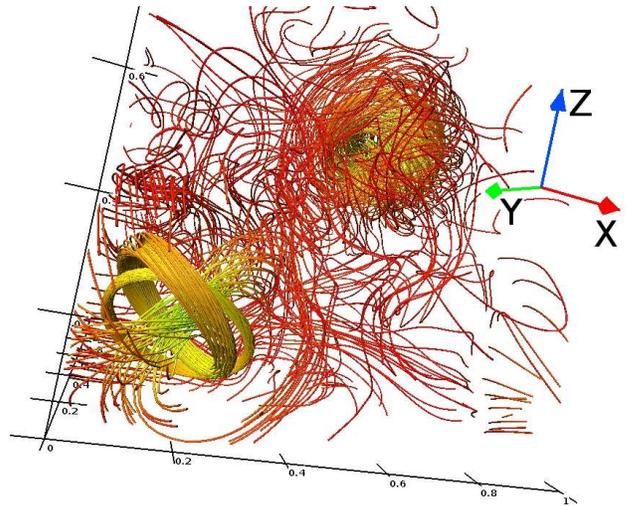}
\end{center}\caption[]{(Color online)
Magnetic field lines at $t=0.248\,t_{\rm res}$ for the
Borromean rings configuration for $\Lu = 1000$.
In the lower-left corner the interlocked flux rings are clearly visible
which differs from the proposed trefoil knot \cite{ruzmaikin:331}.
The flux ring in the opposite corner has an internal twist which
makes it helical.
The colors denote the strength of the field,
where the scale goes from red over green to blue.}
\label{fig: Borromean_256a1_t70}
\end{figure}

Previous calculations showed a significant difference in the decay process
of three interlocked flux rings in the helical and nonhelical case
\cite{fluxRings10}.
In \Fig{fig: energy decay compare} we compare
the magnetic energy decay found from previous calculations
using triple-ring configurations
with the IUCAA knot and the Borromean rings.
The Borromean rings show a similar behavior as the IUCAA knot where the
magnetic energy decays like $t^{-1}$. Similarly to the IUCAA knot we
expect some structure, which is conserved during the relaxation process
and causes the relatively slow energy decay compared to other nonhelical
configurations.
We plot the magnetic field lines at times $t = 0.248\,t_{\rm res}$
and $t = 0.276\,t_{\rm res}$; see
\Figs{fig: Borromean_256a1_t70}{fig: Borromean_256a1_t78}, respectively.
At $t = 0.248\,t_{\rm res}$ there are two interlocked
flux rings in the lower left corner, while in the opposite half of
the simulation domain a clearly twisted flux ring becomes visible.
The interlocked rings reconnect at $t = 0.276\,t_{\rm res}$
and merge into one flux tube with a twist
opposite to the other flux ring. The magnetic helicity stays zero during the
reconnection, but changes locally, which then
imposes a constraint on the magnetic energy decay and could explain
the power law that we see in \Fig{fig: energy decay compare}.
This finding is similar to that of Ruzmaikin and Akhmetiev
\cite{ruzmaikin:331} who propose that, after
reconnection, the Borromean ring configuration transforms first into a trefoil
knot and three 8-form flux tubes and after subsequent reconnection into two
untwisted flux rings (so-called unknots) and six 8-form flux tubes.
We can partly reproduce this behavior, but
instead of a trefoil knot we obtain two interlocked flux rings and, instead
of the 8-form flux tubes, we obtain internal twist in the flux rings.
\begin{figure}[t!]\begin{center}
\includegraphics[width=\columnwidth]{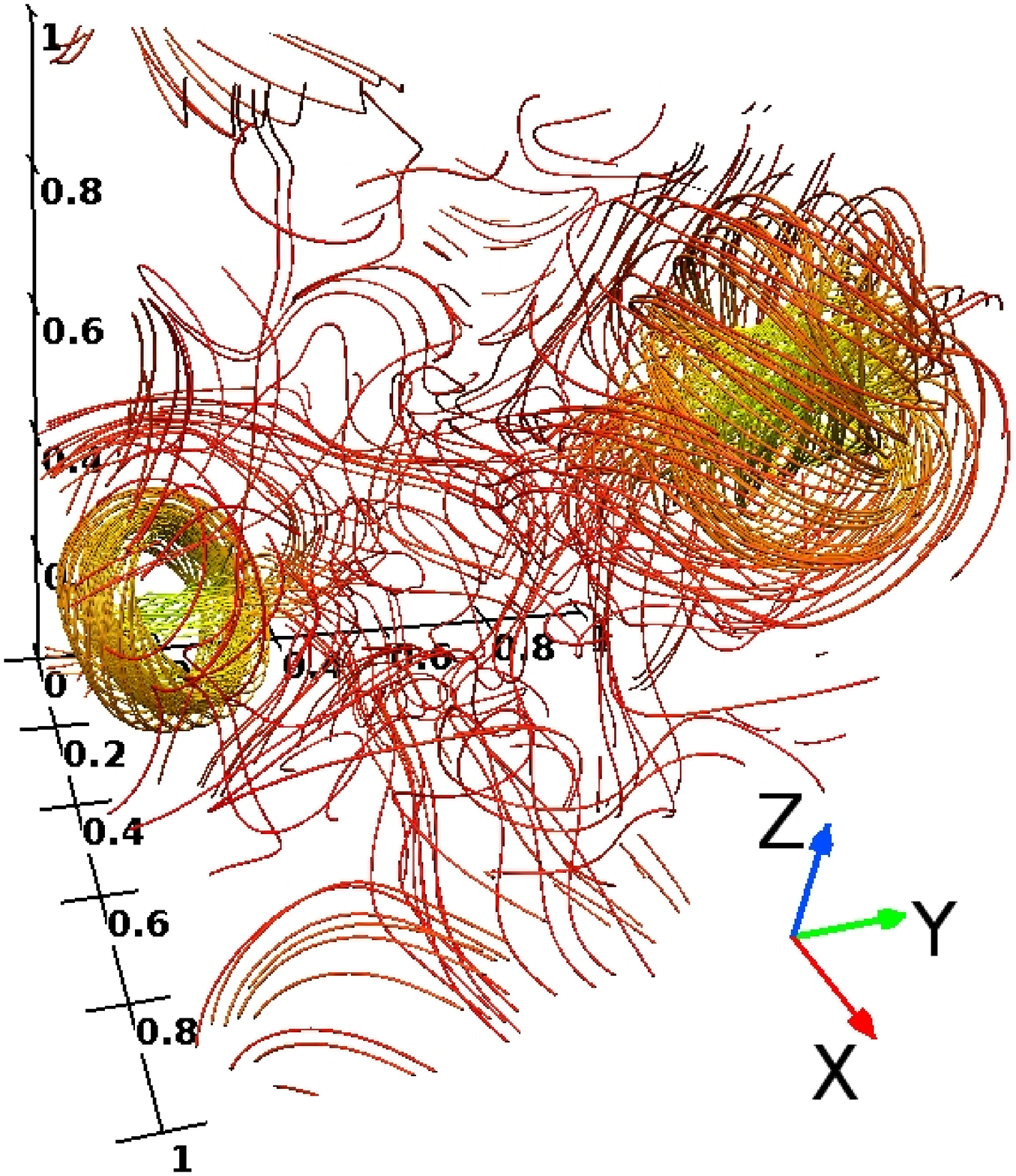}
\end{center}\caption[]{(Color online)
Magnetic field lines at $t=0.276\,t_{\rm res}$ for the
Borromean rings configuration for $\Lu = 1000$.
The two flux rings in the corners both have an internal twist which makes them
helical. The twist is, however, of opposite sign which means that the whole
configuration does not contain magnetic helicity.
The colors denote the strength of the field,
where the scale goes from red over green to blue.}
\label{fig: Borromean_256a1_t78}
\end{figure}

\section{Conclusions}

In this paper we have analyzed for the first time the decay of complex
helical and nonhelical magnetic flux configurations.
A particularly remarkable one is the IUCAA knot for which the
linking number is zero, and nevertheless, some finite magnetic helicity
is gradually emerging from the system on a resistive time scale.
It turns out that both the IUCAA knot and the Borromean rings
develop regions of opposite magnetic
helicity above and below the midplane, so the net magnetic helicity
remains approximately zero.
In that process, any slight imbalance can then lead to the amplification
of the ratio of magnetic helicity to magnetic energy---even though the
magnetic field on the whole is decaying.
This clearly illustrates the potential of nonhelical configurations to
exhibit nontrivial behavior, and thus the need for studying the evolution
of higher order invariants that might capture such processes.

The role of resistivity in producing magnetic helicity from a nonhelical
initial state has recently been emphasized \cite{Low11}, but it remained
puzzling how a resistive decay can increase the topological complexity
of the field, as measured by the magnetic helicity.
Our results now shed some light on this.
Indeed, the initial field in our examples has topological complexity
that is not captured by the magnetic helicity as a quadratic invariant.
This is because of mutual cancellations that can gradually undo themselves
during the resistive decay process, leading thus to finite magnetic
helicity of opposite sign in spatially separated locations.
We recall in this context that the magnetic helicity over the periodic
domains considered here is gauge invariant and should thus agree with
any other definition, including the absolute helicity defined
in Ref.~\cite{Low11}.

Contrary to our own work on a nonhelical interlocked flux
configuration \cite{fluxRings10}, which was reducible to a
single flux ring after mutual annihilation of two rings,
the configurations studied here are non-reducible even
when mutual annihilation is taken into account.

For the helical $n$--foil knot, we have shown that the magnetic
helicity increases quadratically with $n$.
Furthermore, their decay exhibits different power laws
of magnetic energy which lie between $t^{-2/3}$ for the
3--foil knot and $t^{-1/3}$ for the 7--foil knot.
The latter case corresponds well with the
previously discussed case of three interlocked flux rings that
are interlocked in a helical fashion.
The appearance of different power
laws seems surprising since we first expected a uniform power law
in all helical cases in the regime where the magnetic helicity is so large that
the realizability condition plays a role.
This makes us speculate whether there are other quantities that are different
for the various knots and constrain magnetic energy decay.
Such quantities would be higher order topological invariants
\cite{ruzmaikin:331}, which are so far only defined for spatially separated
flux tubes. In order to investigate their role they need to be generalized
such that they can be computed for any magnetic field configuration,
similar to the integral for the magnetic helicity.
    
The power law of $t^{-1}$ in the decay of the magnetic energy for the IUCAA
knot and the Borromean rings is different from the $t^{-3/2}$ behavior
found earlier for the nonhelical triple-ring configuration.
The observed decay rate can be attributed
to the creation of local helical structures that
constrain the decay of the local magnetic field.
But we cannot exclude higher order invariants \cite{ruzmaikin:331} whose
conservation would then constrain the energy decay.

The Borromean rings showed clearly that local helical structures can
be generated without forcing the system. These can then impose constraints
on the field decay. We suggest that spatial variations should be taken into
account to reformulate the realizability condition \eqref{eq: realizability},
which would increase the lower bound for the magnetic energy.
For astrophysical systems local magnetic helicity variations have to
be considered to give a more precise description of both relaxation and
reconnection processes.

Both the IUCAA logo and the Borromean rings do not stay stable during
the simulation time and split up into two separated helical
magnetic structures.
On the other hand we see that the helical $n$--foil knots stay stable.
A similar behavior was seen in \cite{Braithwaite2010MNRAS}, where
magnetic fields in bubbles inside galaxy clusters were simulated.
In the case of a helical initial magnetic field the field decays into a
confined structure, while for sufficiently low initial magnetic helicity,
separated structures of opposite magnetic helicity seem more preferable.

\acknowledgements

We thank the anonymous referee for making useful suggestions
and the
Swedish National Allocations Committee for providing computing resources at the
National Supercomputer Centre in Link\"oping and the Center for
Parallel Computers at the Royal Institute of Technology in Sweden.
This work was supported in part by the Swedish Research Council,
Grant 621-2007-4064, the European Research Council under the
AstroDyn Research Project 227952,
and the National Science Foundation under Grant No.\ NSF PHY05-51164.

\def\araa{Ann. Rev. Astron. Astrophys.}
\def\apjl{Astrophys. J. Lett.}
\def\solphys{Sol. Phys.}
\def\apj{Astrophys. J.}
\def\jgr{J. Geophys. Res.}

\bibliographystyle{ieeetr}
\bibliography{references}
\end{document}